\documentstyle[12pt,epsfig]{article}
\newcommand{\be}[1]{\begin{equation} \label{(#1)}}
\newcommand{\ee}{\end{equation}}
\newcommand{\ba}[1]{\begin{eqnarray} \label{(#1)}}
\newcommand{\ea}{\end{eqnarray}}
\newcommand{\nn}{\nonumber}
\newcommand{\rf}[1]{(\ref{(#1)})}

\def\Lv{$L\hspace{-0.5em}/\ \ $}

\def\znbb{0\nu\beta\beta}
\def\kd{{\rm K}^+\rightarrow \mu^+\mu^+\pi^-}

\def \lg  {\langle}
\def \rg  {\rangle}

\def \znbb {0\nu\beta\beta}
\begin{document}
\hfill{USM-TH-96}\\[1cm]
\begin{center}
   {\Large\bf K-meson neutrinoless double muon decay as a probe of
neutrino masses and mixings}\\[3mm]
Claudio Dib, Vladimir Gribanov$\ ^1$, Sergey Kovalenko
\footnote{On leave from
the Joint Institute for Nuclear Research, Dubna, Russia}
and  Ivan Schmidt\\[1mm]
{\it Departamento de F\'\i sica, Universidad
T\'ecnica Federico Santa Mar\'\i a, Casilla 110-V, Valpara\'\i so, Chile}
\end{center}
\bigskip

\begin{abstract}
Recently an upper bound on the rate of the lepton number
violating decay $K^+\rightarrow \mu^+\mu^+\pi^-$ has been
significantly reduced by the E865 experiment at BNL and further
improvement is expected in the near future.
We study this process as a possible source of information on
neutrino masses and mixings. We find that it is insensitive to the
light(eV domain) and heavy(GeV domain) neutrinos. However due to
the effect of a resonant enhancement this decay is very sensitive
to neutrinos $\nu_j$ in the mass region
$245\mbox{ MeV}\leq m_{\nu_j}\leq 389$ MeV.
At present experimental sensitivity
we deduce new stringent limits on the neutrino
mixing matrix element $U_{\mu j}$ for neutrino masses in this region.
\end{abstract}
\vskip 0.5cm
\section{Introduction}

Nowadays there are sufficient indications to believe that neutrinos are
massive particles mixing with each other \cite{nu-mass}.
These indications come from both experimental
and theoretical sides. The solar neutrinos deficit, the atmospheric neutrino
anomaly and the results of the LSND neutrino oscillation experiment,
all can be explained in terms of neutrino oscillations implying
non-zero neutrino masses and mixings.
On the theoretical side almost all phenomenologically
viable models of the physics beyond the standard model(SM)
predict non-zero masses for neutrinos which can be either
Majorana or Dirac particles.

Majorana masses violate the conservation of total lepton number by
two units $\Delta L = 2$. Therefore lepton number violating(\Lv)
processes represent a most appropriate tool to address the
question of whether neutrinos are Majorana or Dirac particles.
Various  \Lv processes have been studied in the literature in this respect.
Among them there are the neutrinoless nuclear double beta ($\znbb$) decay
\cite{znbb-exp,znbb-rev1},
the decay $\kd$ \cite{Kdec1,Kdec2,Kdec3,Kdec4,NMF},
nuclear muon to positron \cite{DKT} or to antimuon \cite{MMM} conversion,
trimuon production in neutrino-nucleon scattering \cite{tri-mu},
the process
$e^+p\rightarrow \bar\nu l^+_1l^+_2 X$,
relevant for HERA \cite{HERA}, as well as direct production of
heavy Majorana neutrinos at various colliders \cite{heavy-Majorana}.
The analysis made in the literature \cite{Kdec3,lnv} leads
to the conclusion that if these processes are mediated by the Majorana
neutrino exchange then, except $\znbb$-decay, they can hardly
be observed experimentally. This analysis relies on the current
neutrino oscillation data, and on certain assumptions about the neutrino
mass matrix. In the present paper we concentrate on the neutrinoless double
muon decay of kaon $\kd$.
We will show that despite the above conclusion being true
for contributions of the neutrino states much lighter or much heavier
than the typical energy of the $\kd$ decay, there is still a
special window in the neutrino sector which can be
efficiently probed by searching for this process. This window is in
the neutrino mass range
$245\mbox{ MeV}\leq m_{\nu_j}\leq 389$ MeV, where the s-channel neutrino
contribution to the $\kd$ decay is resonantly enhanced, therefore
making this decay very sensitive to the neutrinos in this mass domain.
If neutrinos with masses in this region exist, then from present
experimental data we can
extract stringent limits on their mixing with $\nu_{\mu}$. We
derive these limits from the upper bound on the branching ratio of
$\kd$ recently obtained by E865 experiment at BNL \cite{E865}.

\section{$\kd$ decay in Standard Model with Majorana neutrinos}

In the SM extension with Majorana neutrinos there are two
lowest order diagrams, shown in Fig.1, which contribute to $\kd$ decay.
These diagrams were first considered long ago in Refs.~\cite{Kdec1,Kdec2}.
Here we are studying previously
overlooked aspects of this decay. We concentrate on the s-channel
neutrino exchange diagram in Fig.~1(a) which plays a central role
in our analysis. The t-channel diagram in Fig.~1(b) requires in general
a detailed hadronic structure calculation. In
Ref. \cite{Kdec2} this diagram was evaluated in the Bethe-Salpeter
approach and shown to be an order of magnitude smaller than the
diagram in Fig.~1(a), for light and intermediate mass neutrinos.
As we will see, in the neutrino mass domain of our main interest,
the diagram in Fig.~1(a) absolutely dominates over the t-channel
diagram in Fig.~1(b), independently of hadronic structure.

The contribution from the factorizable s-channel diagram in
Fig.~1(a) can be calculated in a straightforward way, without
referring to any hadronic structure model. A final result for the
$\kd$ decay rate is given by
\ba{rate-Lv}
&&\Gamma(\kd)= c \int\limits_{s_1^-}^{s_1^+} d s_1
\left |\sum_{k} \frac{U_{\mu k}^2 m_{\nu k}}{s_1 - m_{\nu k}^2}\right |^2
G(\frac{s_1}{m_{_K}^2}) + \\ \nn
&&2 \frac{c}{m_{_K}^{2}}{\rm Re}\sum_{k,n}[\int\limits_{s_1^-}^{s_1^+} d s_1
\frac{U_{\mu k}^2 m_{\nu k}}{s_1 - m_{\nu k}^2}
\int\limits_{s_2^-}^{s_2^+} d s_2
\left(\frac{U_{\mu n}^2 m_{\nu n}}{s_2 - m_{\nu n}^2}\right)^*
H(\frac{s_1}{m_{_K}^2},\frac{s_2}{m_{_K}^2})].
\ea
The unitary mixing matrix $U_{ij}$ relates $\nu'_i = U_{ij}\nu_j$
weak $\nu'$ and mass $\nu$ neutrino eigenstates.
The numerical constant in Eq. \rf{rate-Lv} is
$c = (G_F^4/32) (\pi)^{-3} f^2_{\pi}  f^2_{_K} m^5_{_K}
|V_{ud}|^2|V_{us}|^2$,
where $f_{_K} = 1.28 ~f_{_\pi}$, $f_{_\pi} = 0.668~ m_{\pi}$
and $m_K = 494$ MeV is the K-meson mass.
The functions $G(z)$ and $H(z_1,z_2)$ in Eq. \rf{rate-Lv}
after the phase space integration can be written in an explicit
algebraic form
\ba{G-funct}
&&G(z) = \frac{\phi(z)}{z^{2}}
\left[h_{+-}(z)h_{--}(z)-
x_{\pi}^2h_{-+}(z)\right]
\left[x_{\mu}^2 +z -(x_{\mu}^2-z)^2\right]\\ \nn
&&H(z_1,z_2)= h_{--}(z_1)h_{--}(z_2) +
x_{\pi}^2[r_{+}(z_1z_2) - x_{\mu}^2t(z_1,z_2,1)]-r_{-}(z_1z_2)t(z_1,z_2,x_{\mu}).
\ea
Here we defined
$x_{i} = m_i/m_{_K}$
and $h_{\pm\pm}(z)=z\pm x^2_{\pi}\pm x^2_{\mu}$,
$r_{\pm}(z_1z_2)= z_1z_2 -x_{\pi}^2 \pm x_{\mu}^4$,
$t(z_1,z_2,z_3)= z_1+z_2-2 z_3^2$,
$\phi(z)=
\lambda^{1/2}(1,x_{\mu}^2,z)\lambda^{1/2}(z,x_{\mu}^2,x_{\pi}^2)$
with
\mbox{$\lambda(x,y,z) = x^2+y^2+z^2-2xy -2yz- 2xz$.}
The integration limits in Eq. \rf{rate-Lv} are
\ba{lim-int}
&&s_1^- = m_{_K}^2(x_{\pi} + x_{\mu})^2, \ \ \ \
s_1^+ = m_{_K}^2(1 - x_{\mu})^2,\\ \nn
&&s_2^{\pm} = \frac{m_{_K}^2}{2y} \left[2y(1+x_{\mu}^2)-
(1+y-x_{\mu}^2)h_{-+}(y)\pm \phi(y) \right]
\ea
with $y=s_1/m_{_K}^2$.

First we assume that neutrinos can be separated into
light $\nu_k$ and heavy $N_k$ states, with masses
$m_{\nu i} << \sqrt{s_1^-}$  and $\sqrt{s_1^+} << M_{N k}$.
Then the Eq. \rf{rate-Lv} can be approximately written as
\ba{approx-1}
\Gamma(\kd) =
\left|\lg m_{\nu}\rg_{\mu\mu}\right|^2
m_{_K}^{-1} {\cal A}_{\nu}+ \left|\lg M_N^{-1}
\rg_{\mu\mu}\right|^2 m_{_K}^3 {\cal A}_N \\ \nn
- 2\
\mbox{Re}\left[\lg m_{\nu}
\rg_{\mu\mu}\lg M_N^{-1}\rg_{\mu\mu}\right] m_{_K}
{\cal A}_{\nu N},
\ea
where the average neutrino masses are determined in the standard way
\ba{average}
\lg m_{\nu}\rg_{\mu\mu} = \sum_{k=light} U_{\mu k}^2 m_{\nu k},
\ \ \
\lg M_N^{-1}\rg_{\mu\mu}  =  \sum_{k=heavy} U_{\mu k}^2 M_{Nk}^{-1}.
\ea
The approximate formula \rf{approx-1} can be used
for extracting limits on the average neutrino masses
from the experimental data. This leads to the limits:
$|\lg m_{\nu}\rg|\leq Exp({\nu})$, $|\lg M_N^{-1}\rg|\leq Exp({N})$.
We point out that in this case the upper bounds
must satisfy the consistency conditions
$Exp({\nu})<< \sqrt{s_1^{-}}\sim m_{_K}, \
Exp({N})^{-1}>>  \sqrt{s_1^{+}}\sim m_{_K}$.
Otherwise the so derived limits are not applicable to
$\lg m_{\nu}\rg$, $\lg M_N^{-1}\rg $
as it is in Refs. \cite{Kdec4,NMF,HERA}.
If consistency conditions are not satisfied one has to use
the initial formula \rf{rate-Lv}.
Similar consistency conditions take place for
the other \Lv processes.

The dimensionless coefficients in Eq. \rf{approx-1} are
${\cal A}_{\nu} = 4.0\times 10^{-31}, \
{\cal A}_{N} = 7.0\times 10^{-32},  \
{\cal A}_{\nu N} = 1.7\times 10^{-31}$.
With these numbers we can estimate the current upper bound
on the $\kd$ decay rate from the experimental data on other processes.

Atmospheric and solar neutrino oscillation data, combined with
the tritium beta decay endpoint, allow one to set upper bounds on
the masses of the known three neutrinos \cite{barg1}
$m_{e,\mu,\tau}\leq 3$ eV.
Thus in the three neutrino scenario one
gets
$|\lg m_{\nu}\rg_{\mu\mu}| \leq 9\mbox{ eV}$.
In this case we derive from Eq. \rf{approx-1}
the following branching ratio
\ba{limit11}
{\cal R}_{\mu\mu} =
\frac{\Gamma(\kd)}{\Gamma({\rm K}^+\rightarrow all)}
\leq 3.0 \times 10^{-30}  \qquad
{\rm (3\ light\ neutrino\ scenario)}.
\ea
Assuming the existence of heavy ({\it i.e.} order $\sim$ GeV) mass
neutrinos N, we may obtain an upper bound on $\Gamma(\kd)$ using
the current LEP limit on heavy stable neutral leptons
$M_N\geq 39.5$ GeV \cite{LEP}, which leads to
$|\lg M_N^{-1}\rg_{\mu\mu}|\leq n \left(39.5 \mbox{ GeV}\right)^{-1}$,
where $n$ is the number of heavy neutrinos. This limit being substituted
in Eq.~\rf{approx-1} results in the upper bound
\ba{limit12}
{\cal R}_{\mu\mu} \leq 2.0\times 10^{-19} \qquad {\rm
(3\ light +  1\ heavy\ neutrino\ scenario)}.
\ea
Direct searches for $\kd$ decay by E865 experiment
at BNL \cite{E865} give
\ba{kdec-lim}
{\cal R}_{\mu\mu}\leq 3.0\times 10^{-9}
\ \ \ \left(\mbox{ 90\%CL},\ \ \ \mbox{Ref.~}\cite{E865}\right)
\ea
Comparison of this experimental bound with the theoretical
predictions in Eqs.\ \rf{limit11}, \rf{limit12} clearly shows that
both cases are far from being ever detected. On the other hand
experimental observation of $\kd$ decay at larger rates would
indicate some new physics beyond the SM, or, as we will see, the
presence of an extra neutrino state $\nu_j$ with the mass in
the hundred MeV domain. Let us note that such neutrinos are not
excluded by the LEP neutrino counting experiments measuring the
Z-boson invisible width. These experiments set limits not on the
number of light massive neutrinos but on the number of active
neutrino flavors $N_{\nu}=3$. Thus extra massive neutrino states
$\nu_j$ can appear as a result of mixing of the three active
neutrinos with certain number of sterile neutrinos. These massive
neutrinos are searched for in many experiments \cite{PDG}. The
$\nu_j$ states would manifest themselves as peaks in differential
rates of various processes, and can give rise to significant
enhancement of the total rate if their masses lie in an
appropriate region.

\section{Hundred-MeV neutrinos in $\kd$-decay. Resonant case.}

Assume there exists a massive Majorana neutrino $\nu_j$
with the mass $m_j$ in the range
\ba{domain}
\sqrt{s_1^-} \approx 245 \mbox{ MeV} \leq m_{j} \leq \sqrt{s_1^+} \approx 389
\mbox{ MeV}.
\ea
Then the s-channel diagram in Fig.~1(a) blows up because
the integrand of the first term in Eq.~\rf{rate-Lv} has
a non-integrable singularity at $s =m_j^2$. Therefore, in this
resonant domain the total $\nu_j$-neutrino decay width
$\Gamma_{\nu j}$ has to be taken into account. This can be done by
the substitution $m_{j}\rightarrow m_{j} - (i/2)\Gamma_{\nu j}$.
The total decay width $\Gamma_{\nu j}$ of
the Majorana neutrino $\nu_j$ with the mass
in the resonant domain \rf{domain} receives contributions from the following
decay modes:
\ba{channels}
\nu_j\longrightarrow
\left\{\begin{array}{l}
e^+\pi^-,\ e^-\pi^+,\ \mu^+\pi^-,\ \mu^-\pi^+,\\
e^+e^-\nu_e^c,\ e^+\mu^-\nu_{\mu}^c,\ \mu^+e^-\nu_e^c,\ \mu^+\mu^-\nu_{\mu}^c\\
e^-e^+\nu_e,\ e^-\mu^+\nu_{\mu},\ \mu^-e^+\nu_e,\ \mu^-\mu^+\nu_{\mu}.
\end{array}\right. ,
\ea
Since $\nu_j\equiv \nu_j^c$ it can decay in both
$\nu_j\rightarrow l^{-} X(\Delta L=0)$ and
$\nu_j\rightarrow l^{+} X^c(\Delta L=2)$ channels.
Calculating partial decay rates we obtain
\ba{dec-width-4}
\Gamma(\nu_j\rightarrow l \pi)= |U_{lj}|^2 \frac{G_F^2}{4\pi}f_{\pi}^2
m_j^3 F(y_l,y_{\pi})\equiv |U_{lj}|^2 \Gamma_2^{(l)},\\
\Gamma(\nu_j\rightarrow l_1l_2\nu )=
|U_{l_1 j}|^2 \frac{G_F^2}{192\pi^3} m_j^5 H(y_{l1},y_{l2})
\equiv |U_{l_1 j}|^2 \Gamma_3^{l_1l_2},
\ea
where $y_i= m_i/m_j$ and
\ba{kin-fun}
&&F(x,y)= \lambda^{1/2}(1,x^2,y^2)
[(1+x^2)(1+x^2-y^2) - 4 x^2],\\
&&H(x,y)= 12 \int\limits_{z_1}^{z_2} \frac{dz}{z}
(z-y^2)(1+x^2-z)
\lambda^{1/2}(1,z,x^2)\lambda^{1/2}(0,y^2,z).
\ea
The integration limits are $z_1 = y_{l_2}^2,\ z_2=(1-y_{l_1})^2$ and
$F(0,0)=H(0,0)=1$.
Summing up all the decay modes in \rf{channels} one gets for the total
$\nu_j$ width
\ba{total-4}
\nn
&&\Gamma_{\nu j} =
2 |U_{\mu j}|^2(\Gamma_2^{(\mu)} + \Gamma_3^{(\mu e)}+
\Gamma_3^{(\mu\mu)})+
2 |U_{ej}|^2(\Gamma_2^{(e)} + \Gamma_3^{(ee)}+ \Gamma_3^{(e\mu)})\equiv\\
&&\equiv |U_{\mu j}|^2\Gamma_{\nu}^{(\mu)} + |U_{ej}|^2\Gamma_{\nu}^{(e)}.
\ea
In the resonant domain \rf{domain}  $\Gamma_{\nu j}$ reaches
its maximum value at $m_j = \sqrt{s_1^+}$. Assuming for the moment
$|U_{\mu j}|=|U_{ej}|=1$, we estimate this maximum value to be
$\Gamma_{\nu j}\approx 4.7\times 10^{-10}$ MeV.
Since $\Gamma_{\nu j}$ is so small in the resonant domain
\rf{domain} the neutrino propagator in the first term of Eq.~\rf{rate-Lv}
has a very sharp maximum at $s=m_{j}^2$. The second term, being finite
in the limit $\Gamma_{\nu j} =0$, can be neglected
in the considered case. Thus, with a good
precision we obtain from Eq.~\rf{rate-Lv}
\ba{estim3}
\Gamma^{res}(\kd) \approx
c \pi G(z_0)\frac{m_j |U_{\mu j}|^4}
{|U_{\mu j}|^2\Gamma_{\nu}^{(\mu)} + |U_{ej}|^2\Gamma_{\nu}^{(e)}}
\ea
with $z_0 = (m_j/m_K)^2$.
This equation allows one to derive, from the experimental
bound of Eq.~\rf{kdec-lim}, the constraints on $\nu_j$ neutrino mass $ m_{j}$
and the mixing matrix elements
$U_{\mu j}, U_{e j}$ in a form of a 3-dimensional exclusion plot.
However one may reasonably assume that $|U_{\mu j}|\sim |U_{ej}|$.
Then from the experimental bound \rf{kdec-lim} we derive a
2-dimensional $m_j-|U_{\mu j}|^2$ exclusion plot given in Fig.~2.
For comparison we also present in Fig.~2 the existing bounds taken
from \cite{LEP}.
As shown in the figure, the experimental data on
the $\kd$ decay exclude a region unrestricted by the other
experiments. The constraints can be summarized as
\ba{most}
|U_{\mu j}|^2 \leq (5.6\pm 1)\times 10^{-9}\ \ \ \ \
\mbox{for}\ \ \ 245\mbox{ Mev}\leq m_j \leq 385\mbox{ MeV},
\ea
The best limit $|U_{\mu j}|^2 \leq 4.6\times 10^{-9}$ is achieved
at $m_j \approx 300$ MeV. Note that these limits are compatible with our
assumption that $|U_{\mu j}|\sim|U_{ej}|$ since in this mass domain,
typically $|U_{ej}|^2\leq 10^{-9}$ \cite{PDG}.

The constraints from $\kd$ in Fig. 2 and Eq. \rf{most} can be
significantly improved in the near future by the experiments E949 at
BNL and E950 at FNAL \cite{FNAL}.
It is important to notice that in the resonant
domain we have $\Gamma^{res}(\kd)\sim |U|^2$,
while outside $\Gamma(\kd)\sim |U|^4$. Thus in the resonant mass
domain the $\kd$ decay has a significantly better sensitivity to
the neutrino mixing matrix element. In forthcoming experiments
the upper bound on the ratio in Eq.~\rf{kdec-lim} can be improved by
two orders of magnitude or even more.
Then this experimental bound could be translated to
the limit $|U_{\mu j}|^2 < 10^{-11}$ and stronger.

Finally, it is important to mention the possible cosmological
and astrophysical consequences of the hundred-MeV neutrinos considered
in the present paper. Massive neutrinos contribute to the mass
density of the universe, participate in cosmic structure formation,
big-bang nucleosynthesis, supernova explosions, imprint themselves
in the cosmic microwave background etc.\ (see \cite{Raffelt} for a
review). This implies certain constrains on the neutrino masses
and mixings.
Currently, for massive neutrinos in the mass region
of Eq.~\rf{domain}, the only available cosmological
constraints arise from the mass density of the
universe, $\tau_{\nu_j}< (\sim 10^{14})$ sec, and cosmic
structure formation, $ \tau_{\nu_j}< (\sim 10^{7})$ sec,
(taken from ref.~\cite{Raffelt}) where $\tau_{\nu_j}$
is the lifetime of massive neutrinos. On the other hand, on the basis
of the formula \rf{total-4}, assuming $|U_{\mu j}|^2 \sim
|U_{ej}|^2 \leq 4.6\times 10^{-9}$ as in Eq.~\rf{most}, we find
conservatively $10^{-2}$ sec $< \tau_{\nu_j}$. Thus massive
neutrinos with masses in the interval \rf{domain} do not
contradict the known cosmological constraints and there remains a
wide open interval of allowed mixing matrix elements:
$ (\sim 10^{-18}) < |U_{\mu j}|^2,|U_{ej}|^2 < (\sim 10^{-9})$.
Big-bang nucleosynthesis and the SN 1987A neutrino signal may
lead to much more restrictive constraints. Unfortunately, as yet
the analysis \cite{Dolgov} of these constraints does not involve
the mass region \rf{domain}. It may happen that these constraints,
in combination with our constrains in Eq.~\rf{most}, close the
window for neutrinos with masses in the interval \rf{domain}. Then
the only physics left to be studied using the $\kd$ searches would
be physics beyond the SM other than neutrino issues. Nevertheless,
significant model dependence of all the cosmological constraints
should be carefully considered before such a determining
conclusion is finally drawn.

\section{Conclusion}

We studied the potential of the K-meson neutrinoless double muon
decay as a probe of the Majorana neutrino masses and mixings. We found that
this process is very sensitive to the handred-MeV neutrinos $\nu_j$ in the resonant
mass  range \rf{domain}. We analyzed the contribution of these neutrinos
to the $\kd$ decay rate and derived stringent upper limits on the Majorana
neutrino mixing matrix element $|U_{\mu j}|^2$ from the current experimental
data.
In Fig.~2 we presented these limits in the form of
a 2-dimensional exclusion plot, and compared them with existing limits.
The $\kd$ decay excludes a domain previously unrestricted experimentally.
We point out that the current and near future experimental searches
for this decay are not able to provide any information on the
light eV-mass ($m_{\nu}\sim$ eV) or heavy GeV-mass ($m_\nu\sim$ GeV)
neutrinos, since in those cases the required experimental sensitivities are
by many orders of magnitude far from the realistic ones.
Finally, we notice that the decay $\kd$ can in principle probe
lepton number violating interactions beyond the standard model. A
well known example is given by R-parity violating interactions in
supersymmetric models. However these aspects of the $\kd$ decay
are yet to be studied.

\vskip10mm
\centerline{\bf Acknowledgments}
 We thank Ya. Burdanov, O. Espinosa and R. Rosenfeld for helpful discussions.
This work was supported in part by Fondecyt (Chile) under grants
1990806, 1000717, 1980150 and 8000017, and by a C\'atedra
Presidencial (Chile). \bigskip

\newpage
\begin{figure}[h!]
\vspace{-2 cm}
\hspace{-0.5 cm}
\mbox{ \epsfxsize=16 cm\epsffile{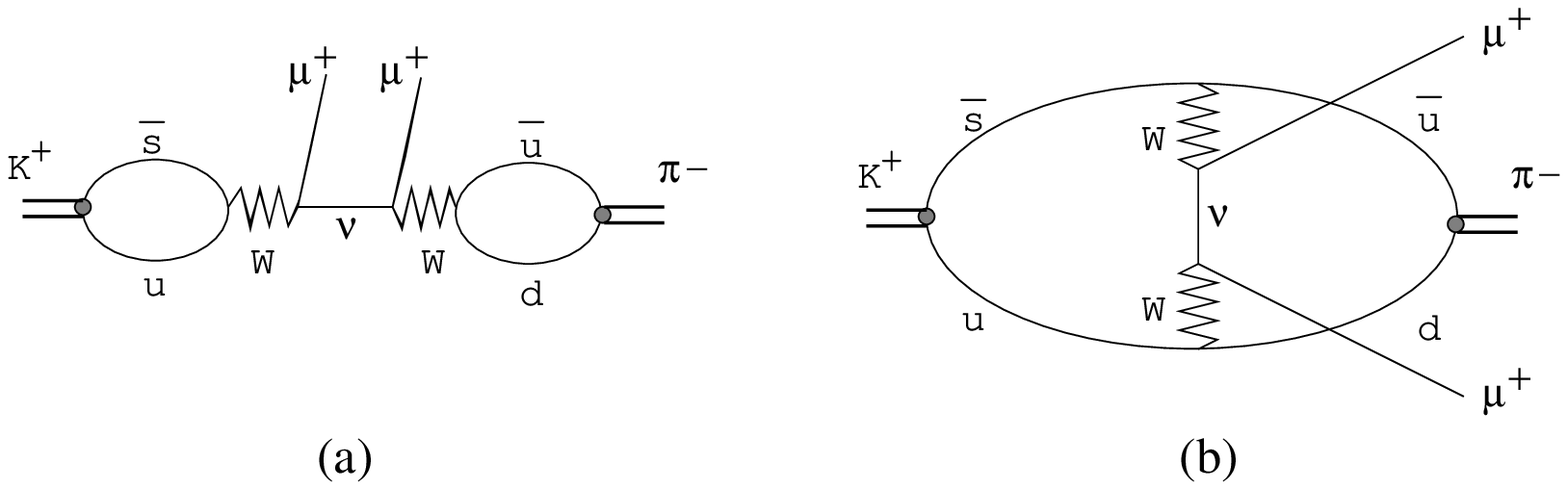}}
\caption{The lowest order diagrams contributing to $\kd$ decay.}
\end{figure}
\begin{figure}[h!]
\hspace{-0.5 cm}
\mbox{\epsfxsize=14 cm\epsffile{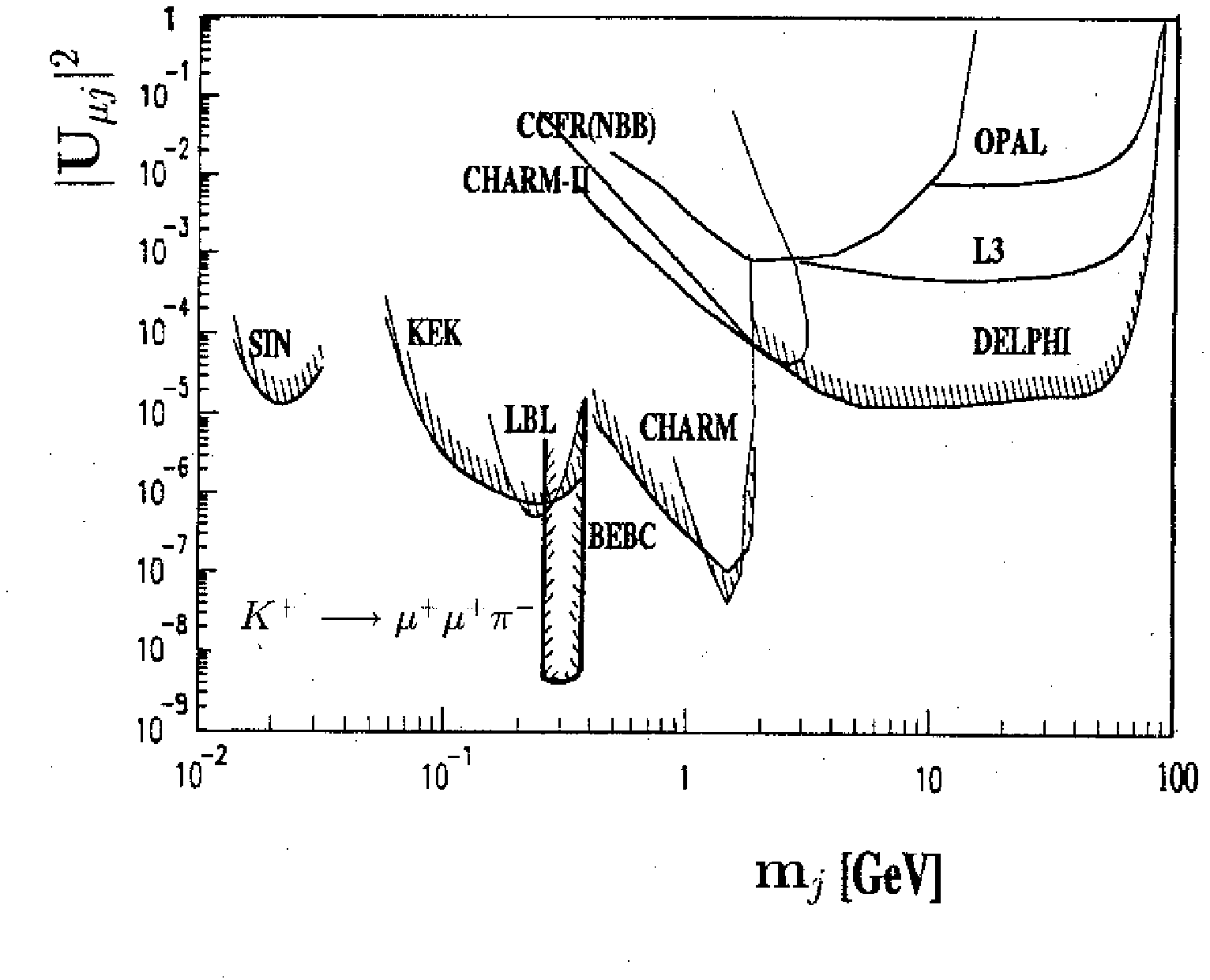}}
\caption{Exclusion plots in the plane $|U_{\mu j}|^2-m_j$.
Here $U_{\mu j}$ and $m_j$ are the heavy neutrino $\nu_j$
mixing matrix element to $\nu_{\mu}$ and its mass respectively.
Domains above the curves are excluded by various experiments
according to the recent update in Ref. \protect\cite{LEP}.
Region excluded by $\kd$ decay [present result]
covers the interval $249$MeV$\leq m_j\leq 385$MeV and extends down to
$|U_{\mu j}|^2\leq 4.6\times 10^{-9}$.}
\end{figure}
\end{document}